\documentclass[fleqn,twoside]{article}
\usepackage{espcrc2}
\newcommand{\adot}{\ensuremath{\dot{\alpha}}}
\newcommand{\bdot}{\ensuremath{\dot{\beta}}}

\title{Lattice Supersymmetry via Twisting}

\author{Simon Catterall\address{Department of Physics\\ 
        Syracuse University, Syracuse NY 13244}%
        \thanks{research supported in part by DOE grant DE-FG02-85ER40231}}
       
\begin{document}

\begin{abstract}
We describe how the usual supercharges of extended supersymmetry may be 
{\it twisted} to produce a BRST-like supercharge $Q$. The usual supersymmetry
algebra is then replaced by a twisted algebra and the action of the twisted
theory is shown to be generically $Q$-exact. In flat space
the twisting procedure can be regarded as a change of variables carrying
no physical significance. However, the twisted
theories can often be transferred to the lattice while preserving the
twisted supersymmetry. As an example we construct
a lattice version of the two-dimensional supersymmetric sigma model.
\vspace{1pc}
\end{abstract}

\maketitle

\section{2D Twisting}
In two dimensions theories with $N=2$ supersymmetry admit a 
global symmetry: $SO(2)\times U(1)_L\times U(1)_R$ where the first factor
reflects rotational invariance (in Euclidean space) with
generator $T$ and the $U(1)$ factors
correspond to chiral symmetries with generators $U_L$ and $U_R$.
Correspondingly the
four supercharges transform as
\begin{eqnarray}
Q_{-+}=(-\frac{1}{2},+1,0)&\;&Q_{--}=(-\frac{1}{2},-1,0)\nonumber\\
Q_{+-}=(+\frac{1}{2},0,+1)&\;&Q_{++}=(+\frac{1}{2},0,-1)
\end{eqnarray}
where the charges are taken in the spin-1/2 representation of
$SO(2)$. The twist consists of decomposing the fields into
representations of a new rotation group \cite{witten}
\[SO(2)^\prime={\rm Diagonal\; subgroup}\left( SO(2)\times SO(2)_{L-R}\right)\]
with new generator $T^\prime=T+\frac{1}{2}(U_L-U_R)$.
The supercharges now transform as
\begin{eqnarray}
Q_{-+}=(0,+1,0)&\;&Q_{--}=(-1,-1,0)\nonumber\\
Q_{+-}=(0,0,+1)&\;&Q_{++}=(+1,0,-1)
\end{eqnarray}
Notice that by this procedure we have produced
two scalar supercharges $Q_L=Q_{-+}$ and $Q_R=Q_{+-}$ and
a spin 1 supercharge with components $Q_{++}$ and $Q_{--}$. Thus we
expect the field theory embodying this twisted supersymmetry will
contain two anticommuting scalar fields and one anticommuting
vector. To match these fields the twisted supermultiplet must also
contain 2 commuting scalars and a commuting vector field.
Furthermore, the original supersymmetry algebra 
\begin{eqnarray}
\{Q_{\alpha +}, Q_{\beta -}\}&=&\gamma^\mu_{\alpha\beta}P_\mu\nonumber\\
\{Q_{\alpha +}, Q_{\beta +}\}&=&\{Q_{\alpha -}, Q_{\beta -}\}=0
\end{eqnarray}
where the spinor index $\alpha ,\beta=+/-$ 
yields a corresponding twisted algebra.
Defining $Q=Q_L+Q_R$ we can see that the latter takes the form
\begin{eqnarray}
Q^2&=&0\nonumber\\
\{Q, Q_{--}\}&=&P_-\;\;\;\{Q,Q_{++}\}=-P_+
\end{eqnarray}
The twisted algebra makes it clear that the momentum operator is $Q$-exact;
that is it may be written as the $Q$-variation of some other operator.
This is a necessary (if not sufficient) condition for
the entire energy-momentum tensor $T_{\mu\nu}$ of 
the twisted theory to be Q-exact. But since 
$T_{\mu\nu}=\frac{\delta S}{\delta g^{\mu\nu}}$ 
$Q$-exactness of $T$ implies that the twisted action is Q-exact too
\[S=Q\Psi(\Phi,\partial\Phi)\]
where $\Psi$ is some function of the fields $\Phi$.
This is a powerful result -- it implies that
the lattice theory obtained by replacing $\Psi$ by an appropriate
lattice function
$\Psi\to\Psi_L(\Phi,\Delta\phi)$ will be exactly invariant under the
twisted SUSY {\it provided only} that I can preserve the nilpotent
property of $Q$ under discretization.
In many cases this can indeed be done
\cite{wz,top,sig}. 

\section{2D Sigma Model}
To make the foregoing arguments more concrete let us construct
the twisted version of the 2D supersymmetric sigma model.
Consider a set of commuting fields $\phi^i(\sigma)$ which act as
coordinates on some
$N$-dim target space with metric $g_{ij}(\phi)$. To ensure
the model has a twisted SUSY we will introduce
scalar grassmann fields $\lambda^i(\sigma)$ and vector 
grassman fields $\eta_{i\mu}(\sigma)$ and another
commuting vector field $B_{i\mu}(\sigma)$. 
We then postulate the following variations of the fields under $Q$
\begin{eqnarray}
Q\phi^i&=&\lambda^i \nonumber \\
Q\lambda^i&=&0 \nonumber \\
Q\eta_{i\mu}&=&\left(B_{i\mu}-
\eta_{j\mu}\Gamma^j_{ik}\lambda^k\right) \nonumber \\
Q B_{i\mu}&=&\left(B_{j\mu}\Gamma^j_{ik}\lambda^k-\frac{1}{2}\eta_{j\mu}
R^j_{ilk}\lambda^l\lambda^k\right)\nonumber
\end{eqnarray}
It is lengthy but straightforward \cite{sig} 
to verify that $Q^2=0$ on all fields. To derive the twisted action consider the
$Q$-exact form
\[
S=\alpha Q\int_\sigma \eta_{i\mu}\left(N^{i\mu}\left(\phi\right)-
\frac{1}{2}g^{ij}B_{j}^{\mu}\right)
\]
If we carry out the $Q$-variation and integrate out $B$ we find
\begin{eqnarray}
S&=&\alpha\int_\sigma
\left(\frac{1}{2}g_{ij}N^{i\mu}N^{j}_{\mu}-\eta_{i\mu}
\nabla_kN^{i\mu}\lambda^k\right.\\
&+&\left.
\frac{1}{4}R_{jlmk}\eta^{j\mu}\eta^{l}_{\mu}\lambda^m\lambda^k\right)
\end{eqnarray}
To generate a kinetic term for the $\phi$ fields it is necessary to
choose
\[
N^{i\mu}=\partial^\mu\phi^i
\]
Finally to make contact with usual sigma model we need to impose
{\it self-duality} conditions on the vector fields in the model. This
amounts to replacing $\eta$ by $P^{(+)}\eta$ where
\[
{P^{i\mu}_{j\nu}}^{\left(+\right)}=
\frac{1}{2}\left(\delta^i_j\delta^\mu_\nu+
J^i_j\epsilon^\mu_\nu\right)
\]
and $J$ is a covariantly constant matrix $\nabla_k J^i_j=0$. Manifolds
possessing such a tensor field are termed K\"{a}hler. It can be shown that the
resultant model can we written in complex coordinates as
\begin{eqnarray}
S&=&\alpha\int d^2\sigma\left(
2h^{+-}g_{I\overline{J}}\partial_+\phi^I
\partial_-\phi^{\overline{J}}\right.\nonumber\\
&-&\left.h^{+-}g_{I\overline{J}}\eta^I_+D_-\lambda^{\overline{J}}-
h^{+-}g_{\overline{I}J}\eta^{\overline{I}}_-D_+\lambda^J\right.\nonumber\\
&+&\left.
\frac{1}{2}h^{+-}R_{I\overline{I} J\overline{J}}
\eta^I_+\eta^{\overline{I}}_-\lambda^J\lambda^{\overline{J}}\right)\nonumber
\end{eqnarray}
Assembling the anticommuting fields into Dirac spinors 
\[\begin{array}{ccc}
\Psi^I=\left(\begin{array}{c}\lambda^{\overline{I}}\\
\frac{1}{2i}\eta^{\overline{I}}_-\end{array}\right)&\;\;\;&
\overline{\Psi}_{I}=\left(\begin{array}{c}\lambda^I\\
\frac{1}{2i}\eta_+^I\end{array}\right)
\end{array}
\]
the quadratic part of the anticommuting action can be written
\[
\overline{\Psi}_{I} \gamma \cdot D\Psi^I
\]
where the Dirac operator in chiral basis is
\[\left(\begin{array}{cc}
0&iD_+\\
-iD_+^\dagger &0
\end{array}\right)\]
Notice that the continuum $Q$-symmetry makes no reference to
derivatives of the fields and so is trivially preserved under discretization.
Thus it appears that we need only replace continuum derivatives
by symmetric finite differences inside $N$ to generate a $Q$-invariant
lattice model with the correct classical continuum limit. Of course
such a procedure will produce both bosonic and fermionic doubles. To remove
these we need to add a Wilson term to the lattice action in such a way
as to preserve supersymmetry. For many K\"{a}hler manifolds
this can be done and the doubles eliminated -- we refer for details
of this construction to Sofiane Ghadab's
talk at this conference.  

\section{4D Twisting}

Theories with $N=2$ SUSY in four dimensions
possess the global symmetry group
\[SU(2)_L\times SU(2)_R\times SU(2)_I\times U(1)\]
The supercharges then transform like
\begin{eqnarray}
Q^i_\alpha &:& (\frac{1}{2},0,\frac{1}{2})\nonumber\\
\overline{Q}_{i\adot} &:& (0,\frac{1}{2},\frac{1}{2})
\end{eqnarray}
As in two dimensions we obtain the twisted theory
by replacing the original symmetry group by $SU(2)_R\times SU(2)^\prime$ where
\[SU(2)^\prime={\rm  Diagonal\; subgroup} \left(SU(2)_L\times SU(2)_I\right)\]
This implies that the isospin index can be treated like a spinor index 
\begin{eqnarray}
Q^i_\alpha&\to& Q^\beta_\alpha\nonumber\\
\overline{Q}_{i\bdot}&\to& G_{\alpha\bdot}
\end{eqnarray}
The trace of $Q^\alpha_\alpha$ is then just our new nilpotent charge
$Q$
and the twisted supercharges now transform as
\begin{eqnarray}
G_{\alpha\bdot}&:&(\frac{1}{2},\frac{1}{2}) --\;{\rm vector}\nonumber\\
Q_{(\alpha\beta)}&:&(1,0) --\;{\rm selfdual\; tensor}\nonumber\\
Q&:&(0,0) --{\rm scalar}
\end{eqnarray}
Corresponding to this we expect the twisted theory to contain an
anticommuting
vector $\psi_\mu$, antisymmetric, self-dual tensor $\chi_{\mu\nu}$ and scalar $\eta$
together with their commuting counterparts.
The original supersymmetry algebra
\[\{Q^i_\alpha,\overline{Q}_{j\bdot}\}=\delta^i_jP_{\alpha\bdot}\]
the becomes the
twisted algebra:
\[\{Q,G_{\alpha\bdot}\}=P_{\alpha\bdot}\; {\rm and}\; \{Q,Q\}=0\]
Notice, once again that the momentum operator $P$ is
again Q-exact! Hence (in most cases) the action is also Q-exact
\[S_{\rm twisted}=Q\Psi\]
Actually the algebra $Q^2=0$ can be generalized in 
a very useful way -- if we are dealing with a gauge
theory we can allow $Q^2={\rm gauge\; transformation}$ without
spoiling the $Q$-invariance of the action. As we will see in our example,
twisted super Yang Mills theories are precisely of this type.
Indeed the twisted supermultiplet contains a set
of commuting fields $A_\mu$,$B_{\mu\nu}$ and $\overline{\phi}$
together with the twisted fermion fields $\psi_\mu$, $\chi_{\mu\nu}$
and $\eta$. All fields
must be taken in the adjoint of some gauge group.
We postulate the following $Q$-transformations \cite{witten2}
\begin{eqnarray}
Q A_\mu&=&\psi_\mu\nonumber\\
Q\psi_\mu&=&iD_\mu\phi\nonumber\\
Q\chi_{\mu\nu}&=&B_{\mu\nu}\nonumber\\
QB_{\mu\nu}&=&[\phi,\chi_{\mu\nu}]\nonumber\\
Q\phi&=&0\nonumber\\
Q\overline{\phi}&=&\eta\nonumber\\
Q\eta&=&[\phi,\overline{\phi}]\nonumber
\end{eqnarray}
Again, it is not hard to show that $Q^2=\delta_G^\phi$ an
infinitesimal gauge transformation with parameter $\phi$.
The $Q$-exact form of the twisted SYM action is just
\[S=Q\int
\frac{1}{4}\eta [\phi,\overline{\phi}]+\chi^{\mu\nu}\left(B_{\mu\nu}+
F^+_{\mu\nu}\right)-i\psi_\mu D_\mu\overline{\phi}\]
In \cite{sugino} Sugino has shown how to generalize the continuum
$Q$-transformations
to the lattice in which the vector field $A_\mu$ is replaced by
a link field $U_\mu$. The resulting theory is gauge invariant,
Q-symmetric and yields the (twisted) super Yang Mills theory in
the naive continuum limit.

Finally, we would like to mention the recent
work by D'Adda et al. \cite{kaw} which points out the intimate
connection between the twisting procedure we have described 
and the representation of fermions in terms of Dirac-K\"{a}hler fields.


\begin{thebibliography}{9}
\bibitem{witten} E. Witten, Comm. Math. Phys. 118:411,1988
\bibitem{wz} S. Catterall and S. Karamov, Phys. Rev. D65 (2002) 094501.
\bibitem{top} S. Catterall, JHEP 0305 (2003) 038
\bibitem{sig} S. Catterall and S. Ghadab, JHEP 0405 (2004) 044.
\bibitem{witten2} E. Witten, Comm. Math. Phys. 117 (1988) 353.
\bibitem{sugino} F. Sugino, JHEP 0403 (2004) 067,JHEP 0401 (2004) 015.
\bibitem{kaw} A. D'Adda, I. Kanamori, N. Kawamoto and K. Nagata,
hep-lat/0406029.
\end{thebibliography}
\end{document}